# Harnessing Structural and Dynamic Heterogeneity to Direct Ion Transport in Plastic Crystal–Polymer Composite Solid-Ion Conductors

Ankit Agrawal[1], Yierpan Aierken[1], Meiling Sun[2], Ethan J. Crumlin[2,3], David Prendergast[1*] and Brett A. Helms[1,4*]

[1] The Molecular Foundry, Lawrence Berkeley National Laboratory, 1 Cyclotron Road, Berkeley, CA 94587.

[2] Advanced Light Source, Lawrence Berkeley National Laboratory, 1 Cyclotron Road, Berkeley, CA 94587.

[3] Chemical Sciences Division, Lawrence Berkeley National Laboratory, 1 Cyclotron Road, Berkeley, CA 94587.

[4] Materials Sciences Division, Lawrence Berkeley National Laboratory, 1 Cyclotron Road, Berkeley, CA 94587.

**Abstract**

Solid-ion conductors (SICs) comprising non-ionic plastic crystals and lithium salts often require compositing with polymers to render them processable for use in solid-state lithium-metal batteries. Here, we show that polymer-doped plastic crystal SICs form a previously unrecognized plastic crystal–polymer high entropy interphase, where ions selectively partition and exhibit a higher fraction of matrix-separated ion pairs than in the bulk. $Li^+$ diffusivity in this interphase is an order of magnitude higher than in other microenvironments due to an increase in the molar volume of the plastic crystal in the vicinity of the polymer, which increases the frequency of bond rotation in the plastic crystal required for ion conduction. Surprisingly, the polymer does not directly participate in ion transport. These insights led us to prepare SICs from specific polymers, plastic crystals, and lithium halide salts that concomitantly deliver fast ion conduction at ambient and sub-ambient temperatures and sustainable passivation of the lithium anode.

**Introduction**

   Plastic crystals are a class of materials that feature weakly interacting organic or organo ionic components arranged in a periodic lattice with long-range order but also short-range disorder, resulting from both conformational and rotational degrees of freedom in the lattice components (*1–5*). Lithium salts can be loaded into a privileged set of plastic crystals to yield soft solid-ion conductors (SICs) for solid-state batteries (*3–8*); in most cases, these soft SICs are composited with polymers to aid in their processing for those devices (*9–11*). The hierarchical complexity inherent to plastic crystal–polymer hybrid SICs has presented significant challenges to building a fundamental understanding their Li$^+$ transport properties. Specifically, it is not known how ions partition or undergo speciation to matrix-separated ion pairs, matrix-shared ion pairs, or contact-ion pairs in various microenvironments embodied in the composite (*9–12*). It is also unclear how matrix, polymer, and ion-solvation dynamics in those microenvironments dictate both ionic carrier concentration and mobility for different combinations of plastic crystalline matrices, lithium salts, and polymers (*9–12*). This gap in knowledge has stunted the further growth of plastic crystal–polymer hybrids as non-combustible SICs for use in Li-metal batteries for the electrification of transportation, particularly for vehicles with large powertrains where increased energy density and passenger safety are of utmost importance and not met with Li-ion batteries in use today (*13–16*).

Here, we use classical molecular dynamics simulations to show that non-ionic plastic crystal–polymer hybrids comprising lithium halide salts in a universal plastic crystal matrix, succinonitrile (SN), form a previously unrecognized plastic crystal–polymer high entropy interphase when composited with specific polymers, such as polyacrylonitrile (PAN). Within this interphase, lithium salts selectively partition and exhibit a higher fraction of matrix-separated ion pairs than in the bulk SN matrix or near the polymer. We find that the diffusivity of Li$^+$ in this interphase is an order of magnitude higher than in other microenvironments and is concomitant with an increase in frequency of bond rotation in SN (i.e., the rate of *trans–gauche* interconversion about the central C–C bond). Empowered by these insights, we prepared several new classes of plastic crystal–polymer SICs to enhance ion conduction below the SIC's melting point, as demonstrated in both LiI- and LiBr-doped SN–PAN composite SICs. Our use of lithium halide salts is unique in plastic crystal–polymer hybrid SICs and builds on prior observations by us and others of the self-passivating character of lithium halides to stabilize Li metal–electrolyte interfaces (*17–21*). We further demonstrate that only three monolayers of lithium halides are necessary at the interface to prevent electropolymerization of SN, which substantially raises the prospects of using SN-based SICs in Li-metal batteries (*6, 22, 23*). We also comprehensively assess the limiting current density, Sand time and capacity (*24, 25*), and evolution of area specific resistance (*ASR*) for these SICs in

Li-metal cells and show that halide redox can be used to protect Li metal cells from overcharging (*26*).

Our work brings into focus with molecular precision the foundational mechanisms for Li$^+$ transport in plastic crystal–polymer composite SICs. Specifically, bulk transport measurements and dynamic information from nuclear magnetic resonance (NMR) spectroscopy or broadband dielectric spectroscopy for plastic crystal–polymer composite SICs have led to an assumption that enhancements in Li$^+$ conductivity are due to faster Li$^+$ transport along the polymer backbone (and with specific functionality along the polymer backbone) rather than in the bulk of the SN matrix (*10*). This assumption is difficult to reconcile with the mismatch between comparably faster conformational dynamics of *trans–gauche* interconversion in the bulk plastic crystal relative to slower polymer segmental chain dynamics. This assumption is further challenged by unexplained and highly contrasting behaviors by different polymers in Li$^+$-doped SN–polymer SICs, where in some cases Li$^+$ transport is enhanced and in other cases it is impaired, both with respect to overall conductivity and cation transference number (*9–11*). Thus, a molecular understanding of how specific polymers "dope" plastic crystal–polymer composite SICs to enhance Li$^+$ transport has not been forthcoming until our work. We also resolve through our use of self-passivating lithium halide dopants the outstanding problem of SN electropolymerization at lithium metal and provide new insights into the dendrite-suppressing character of lithium halide-doped plastic crystal–polymer hybrid SICs in Li-metal cells (*3, 6–8, 22, 23, 27*).

**Results**

SICs comprising polymer–plastic crystal composites loaded with lithium salts presents formidable challenges to studying structure–transport relationships at the molecular level due to ambiguity in solvation structure and mobility of Li$^+$ in various microenvironments in these hierarchically complex materials. To begin unraveling these details, we used classical molecular dynamics (MD) simulations (*28*) to differentiate various solid-solvation environments for Li$^+$ and quantify time-dependent Li$^+$ displacement therein for SN matrices doped with 0.10 M LiI, either in the presence or absence of PAN, the polymer analogue of SN. Notably, the equilibrated structure of LiI-doped SN–PAN (Figure 1A) exhibited the full spectrum of expected behavior: LiI speciation as matrix-separated ion pairs, matrix-shared ion pairs, or contact-ion pairs; as well as significant structural distortions in the SN lattice in proximity to PAN relative to that observed for LiI-doped SN without the polymer (Figures 1A & 1B). We also found that by introducing PAN to LiI-doped SN SICs, the aggregate diffusion coefficient ($D$) for Li$^+$ increased 4-fold—from $0.358 \times 10^{-7}$ cm$^2$ s$^{-1}$ to $1.422 \times 10^{-7}$ cm$^2$ s$^{-1}$ (Figure 1C).

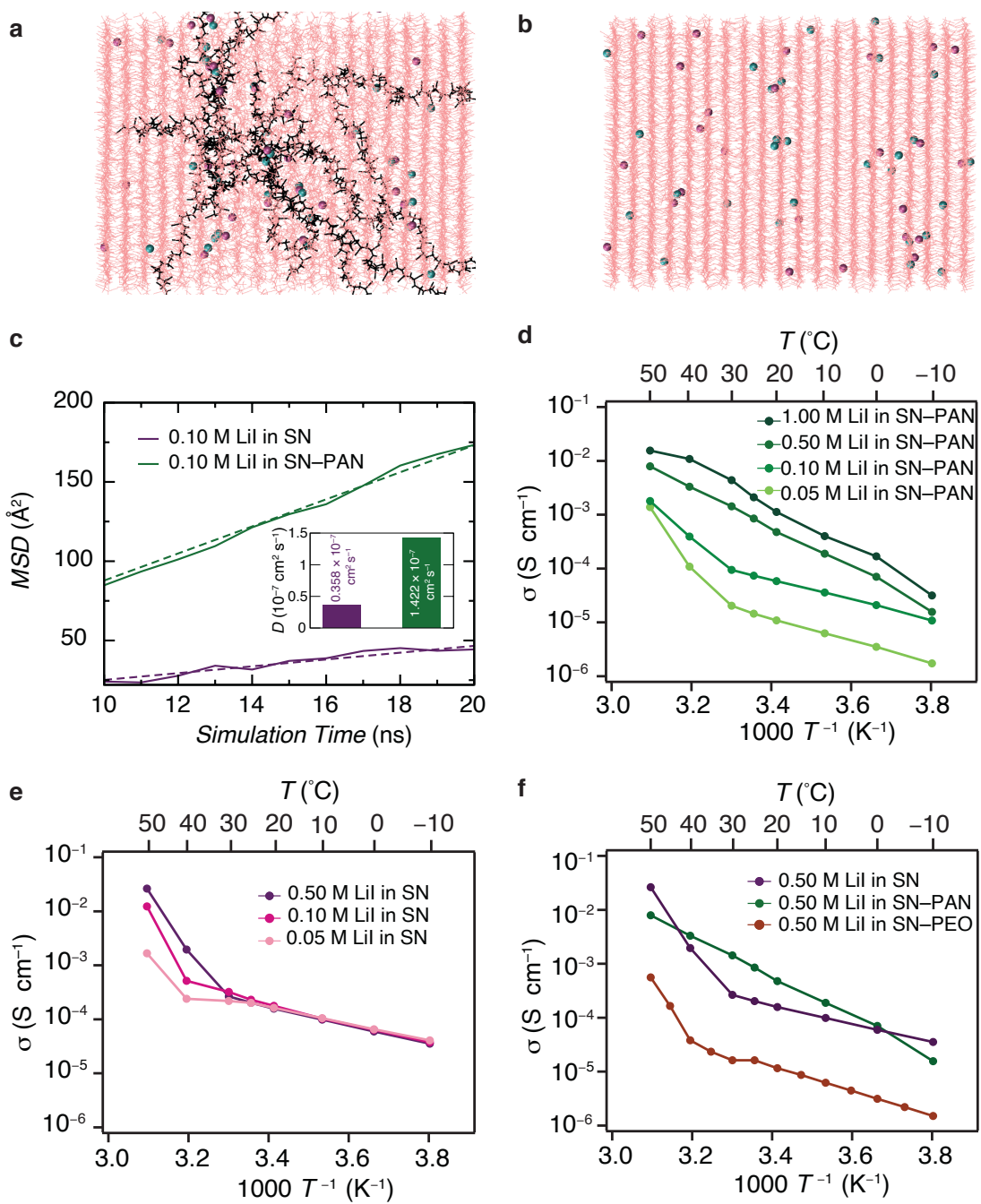

**Figure 1: Lithium transport through SIC composites.** (A) Simulated structure of a LiI-doped SN–PAN plastic crystal–polymer composite, showing SN molecules (light pink), 13 linear PAN chains (black), 0.10 M of Li$^+$ (cyan) and I$^-$ (pink). (B) Simulated structure of LiI-doped SN plastic crystal SIC. (C) Simulated mean square displacement (*MSD*) of Li$^+$ as a function simulation time

within LiI-doped SN–PAN and SN-only SICs. (D) Temperature-dependent ionic conductivity of LiI-doped SN–PAN composite SICs with [LiI] = 0.05 M, 0.10 M, 0.50 M and 1.00 M (light green to dark green). (E) Temperature-dependent ionic conductivity of LiI-doped SN, where [LiI] = 0.05 M, 0.10 M, and 0.50 M (light purple to dark purple). (F) Temperature-dependent ionic conductivity of 0.50 M LiI-doped SN, 0.50 M LiI-doped SN–PAN composite, and 0.50 M LiI-doped SN–PEO composite.

Inspired by these computational studies, we prepared a series of LiI-doped SN–PAN composite SICs by dissolving LiI (0.05–1.00 M) and PAN (7.5 wt%, $M_w$ = 150,000 g mol$^{-1}$) in molten SN at 80 °C for a period of 24 h (Figure S1 and S2); similarly, we prepared LiI-doped SN (0.05–0.50 M) and LiI-doped SN–PEO (0.50 M) to serve as controls. We determined the ionic conductivity of 150-μm thick samples of each SIC placed between stainless steel blocking electrodes by electrochemical impedance spectroscopy (EIS) and fitting the spectra to an equivalent circuit (Figure S3). The ionic conductivity at 25 °C ($\sigma_{25C}$) for SN–PAN composites was well-behaved, increasing monotonically from 0.05 to 5.0 mS cm$^{-1}$ for a LiI doping range of 0.05–1.00 M (Figure 1D). In stark contrast, $\sigma_{25C}$ for SN was largely invariant (0.10 mS cm$^{-1}$) to increases in LiI doping, which was limited to ~0.50 M due solid-solvation limits of SN (i.e., half of what is possible when PAN is present) (Figure 1E). Notably, $\sigma_{25C}$ increased 50-fold for SN–PAN composite SICs relative to those containing only SN and LiI. This behavior was also found for composites and matrices doped with LiBr instead of LiI (Figure S4). We also found that PAN was a key contributor to the observed increase in $\sigma_{25C}$. For example, composite SICs prepared with polyethylene oxide (PEO) as polymer inclusions exhibited $\sigma_{25C}$ of only 10$^{-5}$ S cm$^{-1}$ (Figure 1F), which is an order of magnitude lower than LiI-doped SN and over two orders of magnitude lower than LiI-doped SN–PAN composites. Not surprisingly, once molten, the ionic conductivities of lithium halide-doped SN matrices compare more favorably to lithium halide-doped SN–PAN composites.

The enhanced ionic conductivity of LiI-doped SN–PAN composites at near-ambient temperatures is significant and strongly depends on the degree to which both lithium salts as well as the polymer co-dope the composite. Powder X-ray diffraction (PXRD) analysis of LiI-doped SN SICs in the presence and absence of PAN polymer inclusions (Figure S5), exhibited (0 1 1) and (0 2 2) Bragg peaks of crystalline SN shifting to lower 2θ with increasing LiI loading up to 0.10 M, signifying an expansion of the SN lattice. However, above 0.10 M LiI doping in the SN–

PAN composite, the (0 1 1) and (0 2 2) peaks split, indicating the formation of distinct SN phases within the SIC composite (Figure S5B–S5C).

The emergence of heterogeneous phase behavior in the SN lattice concomitant with increases in ionic conductivity warranted deeper investigation into the structural and dynamic underpinnings for $Li^+$ transport in LiI-doped SN–PAN composites. Returning to our MD simulations, we examined the spatial distribution and speciation of LiI in both SN and SN–PAN composite SICs and characterized the ion solvation dynamics with spatial precision. Specifically, we determined the radial distribution function, $g(r)$, for $Li^+$ relative to its distance from the polymer backbone. From the valleys in $g(r)$, we identified three distinct domains within the LiI-doped SN–PAN composite: 1) $Li^+$ within 3.5 Å (brown zone) of the PAN chain; 2) $Li^+$ between 3.5–5.5 Å (green zone) from PAN, which we denote as the interphase region (with distinct characteristics defined in what follows); and 3) $Li^+$ beyond 5.5 Å (purple zone) from PAN, within a bulk-like SN phase (Figure 2A–B). When comparing $Li^+$ diffusivity for different zones, we noted highly contrasting behaviors (Figure 2C). In the immediate vicinity of PAN chains (brown zone), $Li^+$ are persistently bound to nitrile pendants along the polymer backbone and their motion is correlated with that of PAN, which has a low diffusion constant. Far away from PAN (purple zone), the microenvironment resembles bulk SN and $D$ for $Li^+$ is also low. Notably, however, in the interphase (green zone), $D$ for $Li^+$ is an order of magnitude higher than the two phases bounding it. The interphase is distant enough from the PAN backbone that $Li^+$ are surrounded only by SN molecules, rather than polymer chains; the partial coordination number of $Li^+$ to $I^-$ within the interphase is characteristically low, as evidenced by plotting the radial distribution function of $Li^+$ with respect to distance from $I^-$ (Figure S6).

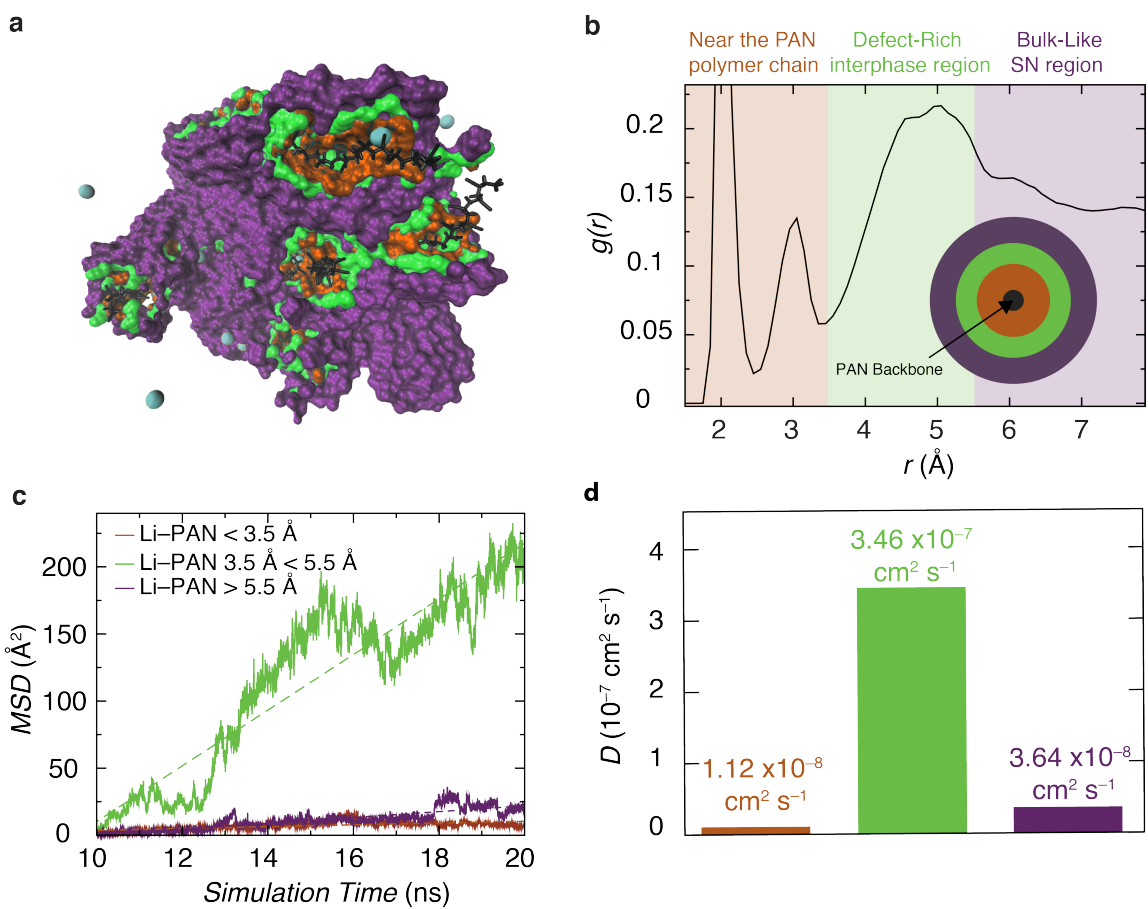

**Figure 2 Different microenviroments within the SIC composite**. (A, B) Representation of three domains based of the radial distribution function, $g(r)$, of $Li^+$ concentration with respect to the distance from the PAN backbone: within 3.5 Å (brown), between 3.5 Å and 5.5 Å (green), and more than 5.5 Å (purple) from the PAN backbone (black). (C) The simulated mean square displacement (*MSD*) of $Li^+$ as a function of simulation time within different domains inside LiI-doped SN–PAN composite. (D) $Li^+$ diffusion coefficient (*D*) calculated from the slope of the *MSD*.

To understand the mechanism of ion transport in the interphase, we analyzed the mean square displacement (*MSD*) of SN, $Li^+$, and $I^-$ for 20 ns in the MD simulations. Both $Li^+$ and $I^-$ displacements were an order of magnitude higher than that for SN molecules in the composite; this observation does not change when considering ion dynamics in SN without PAN (Figure S7). Consequently, $Li^+$ does not undergo vehicular diffusion in either SN or SN–PAN matrices, as is common for liquid electrolytes with coordinating organic solvents. On the contrary, $Li^+$ undergoes structural diffusion, i.e., correlated with the rotation of nitrile groups around the C–C bond in both coordinating and neighboring SN molecules (*1, 3, 8*). Specifically, as SN molecules undergo *trans–gauche* interconversion about the central C–C bond (orange colored SN in Figure 3A and 3B), perturbation in local solid-solvation environments leads to the hopping of Li$^+$ from the initial solvation site (Figure 3A, left) to a neighboring site (Figure 3A, right).

Given the heterogeneity in solid-solvation environments for Li$^+$ in SN–PAN composites as well as the faster diffusivity of Li$^+$ in the interphase, it follows that the dynamics of bond rotation in SN in the interphase should be characteristically faster than in other regions of the SIC. To differentiate the frequency of rotation of the central C–C bond in SN within the interphase and nearby bulk-like SN, we randomly chose a group of SN molecules within each domain and followed N–N distances with time. The N–N distances were used to characterize each molecule as *gauche* (3.71 Å) or *trans* (5.59 Å) (Figure 3B). SN molecules in the interphase exhibited more frequent interconversion events. SN situated farther from PAN were, on the other hand, arrested in the *gauche* conformation and rarely interconverted to the *trans* conformer. The higher rate of *trans–gauche* interconversion in the interphase region was correlated with faster Li$^+$ mobility through the SIC (Figure 3B).

Consistent with the peak shifts to lower angles in our XRD data (Figure S5), our simulations reveal higher molar volumes for SN molecules within the interphase (Figure 3C). This increase in local free volume results in weaker intermolecular interactions between neighboring SN molecules and facilitates comparably faster conformational dynamics. Our observations establish a direct correlation between the increased frequency of bond rotation in SN and Li$^+$ diffusion through a spatially defined high entropy interphase within the SIC. We conclude, therefore, that PAN inclusions in SN matrices induce the formation of a ~2-Å thick interphase ~3.5–5.5 Å away from the polymer backbone that is enriched with matrix-separated ion pairs. The larger molar volume for SN within the interphase results in greater freedom for SN molecules to undergo *gauche–trans* interconversion, which assists the ion transport process, increasing the diffusivity of Li$^+$ ions.

This level of precision in linking macroscale transport properties to local Li$^+$ solid-solvation structures and SN dynamics in an emergent high entropy interphase has not been forthcoming in previous work on plastic crystal–polymer composite SICs and, importantly, resolves an outstanding contradiction in the field. Our analysis indicates that while the polymer is responsible for inducing the formation of a high entropy interphase, the polymer does not directly participate in Li$^+$ transport driving the ion current when an electrochemical cell is polarized. Furthermore, the suppression of *gauche–trans* interconversion events in bulk SN matrices may fundamentally limit the use of SN-based SICs in electrochemical cells operating at near-ambient temperatures, whereas with PAN–SN composites, this constraint is relaxed substantially due to faster ion transport within interphase regions that percolate throughout the composite.

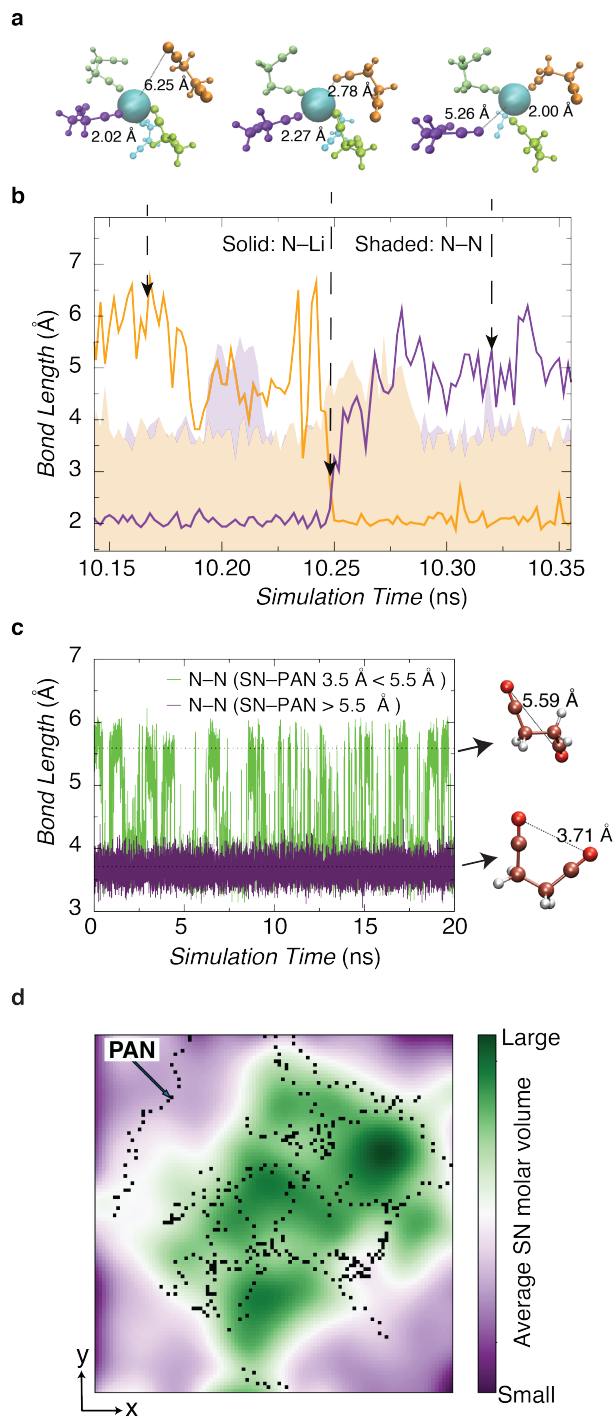

**Figure 3 Li⁺ and SN displacement dynamics within the SIC** (A–B) The simulated bond length between the Li⁺ and SN within the first and second solvation shell (solid lines), as well as intra N–N distances for different SN molecules (shaded zone). The solvation shell of Li⁺ at different simulation times shows Li⁺ hopping between adjacent solvation sites assisted by the *trans* to

*gauche* conformational change of neighboring SN molecules (orange shade) (C) Rotational dynamics for *trans–gauche* interconversion within the high entropy interphase (green) and nearby bulk-like SN (purple). (D) Color map depicting the average SN molar volume, obtained from Voronoi analysis (*29*), averaged along the *z*–direction.

The co-doping of plastic crystal SICs with lithium halide salts and PAN polymer chains, while beneficial to ion transport, does not directly address the known reductive instability of SN at anode–electrolyte interfaces in solid-state Li metal cells and precludes their use as such (*6, 22, 23, 30*). Our MD simulations and density functional theory (DFT) calculations using the Vienna *ab initio* simulation package (VASP) (*31–34*) show that even one molecular layer of SN sandwiched between two Li metal surfaces (Figure 4A) causes SN to undergo electropolymerization to form an insulating layer on the anode (Figure 4B–C). We hypothesized that the self-passivating character of lithium halides at Li–electrolyte interfaces (*17–20, 35*) could be leveraged for stabilizing SN-based SICs against electropolymerization, which has not been demonstrated previously. To test the efficacy of LiI as an electron-blocking layer, we devised an electronic structure simulation cell consisting of a single molecule layer of SN separated from the Li metal anode by three atomic layers of LiI (Figure 4D). After structure relaxation, both LiI and SN layers preserve crystalline structure, and no electropolymerization is observed (Figure 4D). To further understand the electronic hybridization at the interface between LiI and Li metal, we used Bader charge analysis to calculate the net charge distribution within the simulation cell (*36–39*). From the net atomic charge distribution across the Li metal and LiI, we were able to confirm the presence of an electrical double layer: in particular, $I^-$ anions near Li metal atoms raise the local chemical potential for electrons, blocking them from transferring from Li metal to SN molecules (Figure S9). This was also confirmed from the electronic density of states (DOS), where we observed charge transfer and partial reduction of iodide anions at the anode–SIC interface, blocking further electron transfer from the anode to SN molecules.

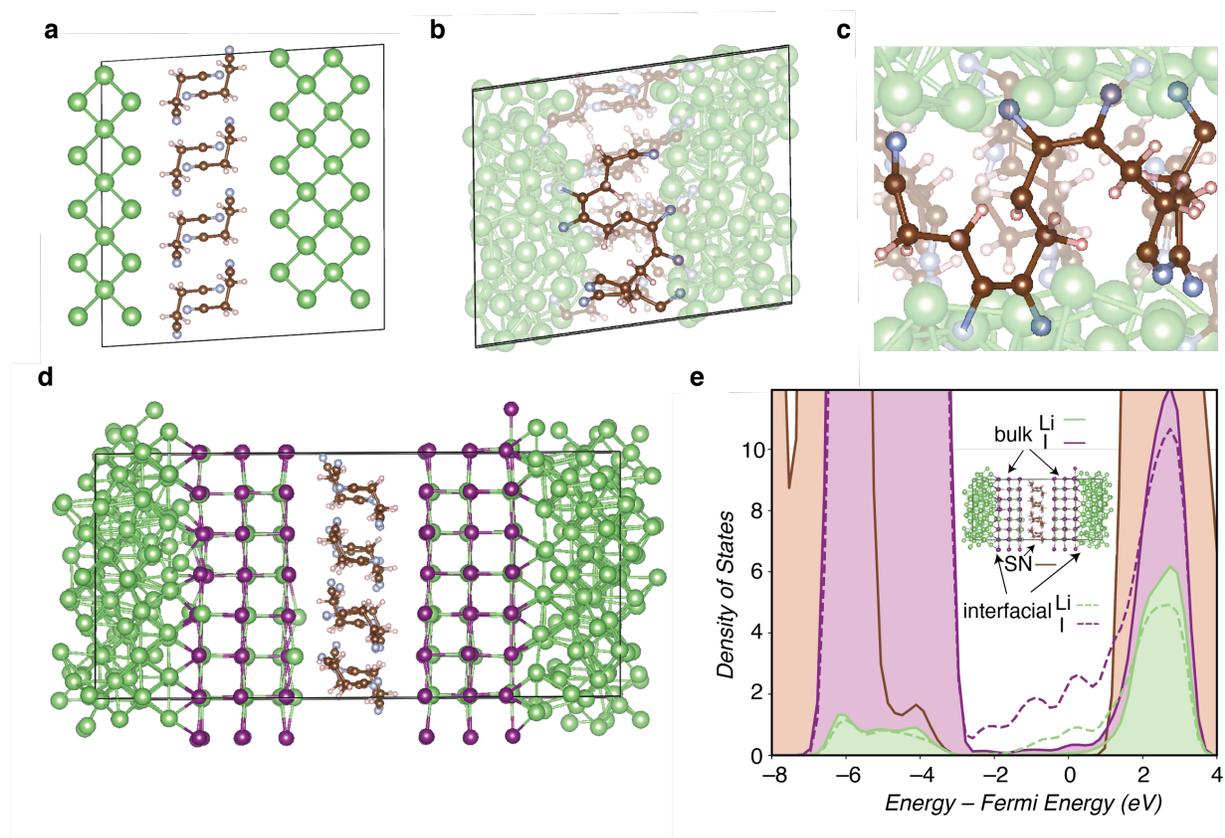

**Figure 4 Understanding and controlling the electrochemical stability of SN against lithium metal.** (A) Initial state of the simulation cell containing SN molecules sandwiched between Li metal (green). (B,C) The relaxed state of the simulation cell shows the electropolymerization of SN when SN comes in contact with lithium metal. (D) The relaxed state of simulation cell where LiI separates Li metal and SN while preventing electropolymerization of SN. (E) The density of states of SN between LiI–coated Li metal.

To test the efficacy of LiI as an electronically insulating and ionically conductive interlayer, we galvanostatically cycled Li–Li symmetric cells assembled with 150-μm thick SN–PAN composite SICs loaded with 0.5 M LiI; the Li electrodes were pre-coated with LiI. At 25 °C, the initial overpotential required for Li plating/stripping was 15 and 30 mV vs. Li/Li$^+$ at current densities of 25 and 50 μA cm$^{-2}$, respectively (Figure 5 A–D). In contrast, without any stabilizing interlayer (negative control) the initial plating over potential was more than 10–fold higher, at around 150 mV vs. Li/Li$^+$, due, presumably, to the formation of an insulating SEI (Figure S10). The area-specific resistance (*ASR*) of Li–Li symmetric cells featuring LiI interlayers at steady-state was ~60 Ohm-cm$^2$ at 50 µA cm$^{-2}$ and showed scant growth after 100 cycles of plating/de–plating, where 50 µAh cm$^{-2}$ of Li was plated in each cycle (Figure S10). Without a stabilizing layer, and following 10 cycles of plating/de-plating, ASR grows as high as 1000 Ohm-cm$^2$ at 25 µA cm$^{-2}$. Due to the high ionic conductivity and low interfacial impedance in our symmetric cells, ASR values at 25 °C are lower by at least an order of magnitude than previously reported non-ionic plastic crystal SICs (*23*), where cell resistances of 220 Ohm increased to 450 Ohm within 100 cycles of plating/deplating at 50 µA cm$^{-2}$ for 100 µA cm$^{-2}$ of Li plated in each cycle. They also compare favorable to ionic plastic crystal SICs, which typically feature lower ionic conductivity (< 10$^{-4}$ S cm$^{-1}$) at room temperature and therefore give rise to higher ASR (*40–42*). For example, using diethyl(methyl)(isobutyl)phosphonium hexafluorophosphate ([P$_{122i4}$][PF$_6$]) as the matrix, the area-specific resistance (*ASR*) of Li–Li symmetric cells was around 200 Ohm-cm$^2$ at 40 °C while plating at 100 µA cm$^{-2}$ for 100 µAh cm$^{-2}$ of Li plated in each cycle (*40*).

Given the stability of Li–Li symmetric cells with LiI interlayers, we were further able to determine the cation transference number at steady-state ($t_{+ss}$ = 0.72) for LiI-doped SN–PAN SIC as well as the electrochemical stability window by linear sweep voltammetry (LSV) (Figures 5E & 5F). The relatively high value for $t_{+ss}$ is consistent with our assessments of the solvation structure and transport mechanism for Li$^+$ in the composite via structural diffusion; for comparison, $t_{+ss}$ for PEO electrolytes range from 0.1–0.2 depending on the formulation (*43*). The high-voltage stability limit of lithium halide-doped SN–PAN composites is dictated by anion redox, which is tunable for different halide salts (*44*).

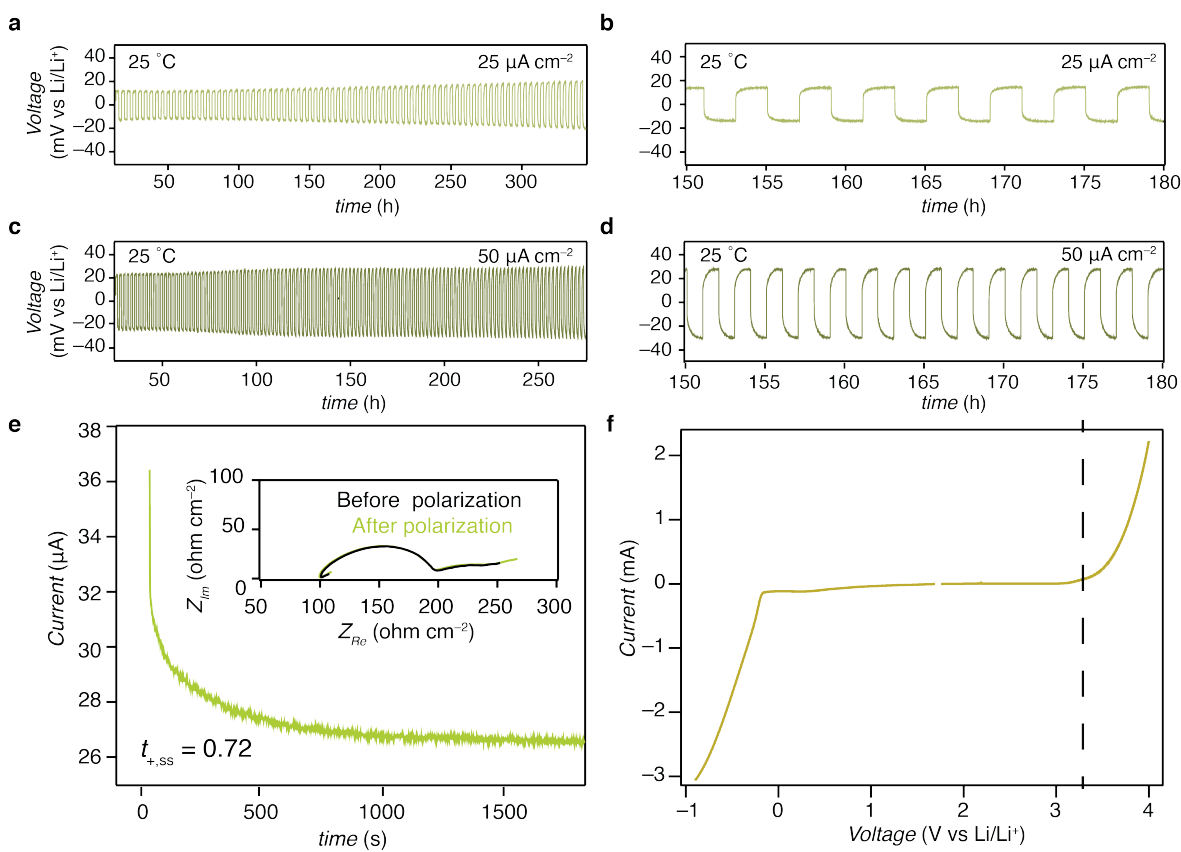

**Figure 5 Electrochemical characterization of the SIC composite**. (A–D) Reversible Li plating and stripping in LiI-doped SN–PAN SIC at the current density of 25 µA cm$^{-2}$ (A–B, light green) and 50 µA cm$^{-2}$ (C–D, dark green) (E) Current transient of a Li|SIC|Li cell under 10 mV polarization. Inset: impedance response of the symmetric cell before and after polarization shows an only minuscule change. The ratio between the steady-state current and peak current relates the steady-state cation transference number, $t_{+\text{ss}}$. (F) Current response of linear sweep voltammetry at the scan rate of 10 mV s$^{-1}$ across the Li|SIC|Cu cell. Dashed line indicates the onset of the SIC oxidation, which is dictated by halide redox.

To evaluate the safe operating limits for LiI-doped SN–PAN SICs, we plated Li metal from one electrode to the other galvanostatically in Li–Li symmetric cells whose anodes were each configured with the protective LiI interlayer. We found that at current densities less than 250 µA cm$^{-2}$ (i.e., largest sustainable current density), the plating overpotentials reached steady-state. Above 500 µA cm$^{-2}$ (i.e., smallest unsustainable current density), however, the overpotential increased slowly over time. We calculated the limiting current as the average of the largest sustainable current and smallest unsustainable current (375 µA cm$^{-2}$) (*45, 46*).

The continuous growth of an insulating layer at the interface also manifests as destabilizing for long operating times; the underlying mechanical instability at the anode–electrolyte interface may result in a spike in plating/stripping overpotential (*19, 47–49*). During extended plating times, both below and above the limiting current, we observed step-changes to higher yet constant overpotential ~3.1 V (Figure 6B and 6C), which is not typically observed in SICs yet consistent with anion redox noted in the LSV (Figure 5F). Thus, while at early stages of Li metal plating, transport limitations at current density above the limiting current manifest as the primary determinant of instability in the cells, over extended plating periods, even below the limiting current, interfacial instabilities begin to take hold. However, regardless of the current density, when step-changes are observed, there are no ensuing signs of internal shorts from dendrites (*45, 46*) across a wide range of current densities and operating temperatures. Halide redox therefore emerges as an intrinsic feature in lithium halide-doped SN–PAN composites that can be leveraged as protection against overcharging in Li metal cells (Figure 6 B–C).

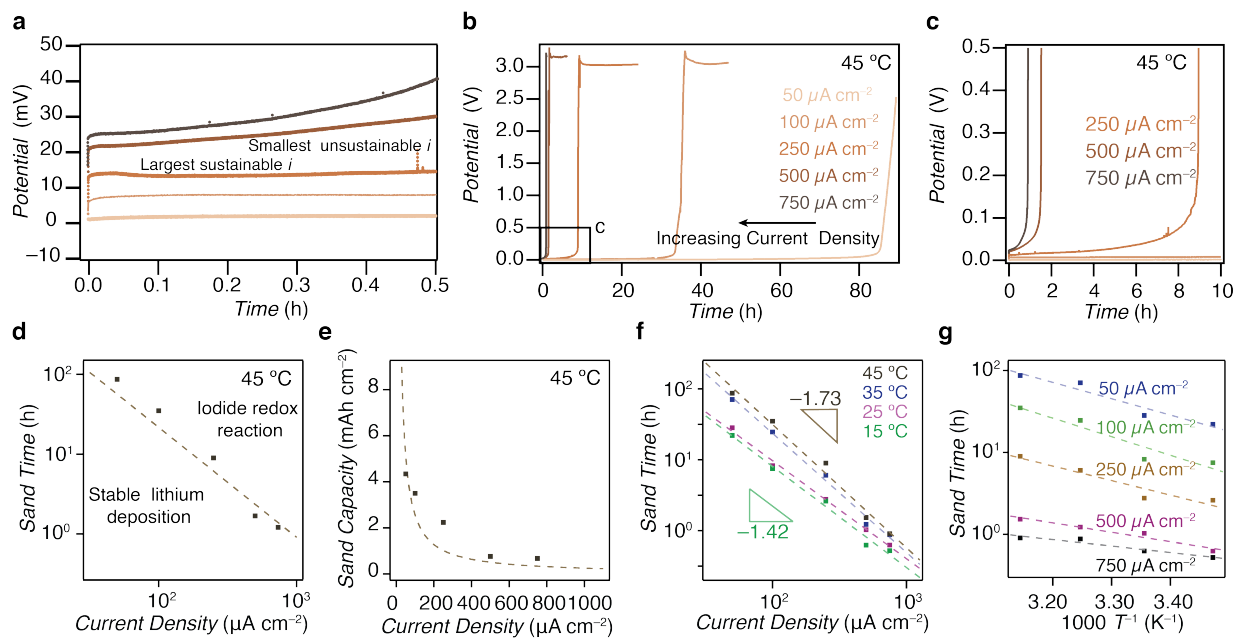

**Figure 6 Sand Analysis of the SIC composite.** (A) Time-dependent potential behavior of 150 µm thick 1.00 M LiI-doped SN–PAN composite in response to applied current densities (50 to 750 µA cm$^{-2}$, light brown to dark down) for first 30 min at 45 °C. The largest sustainable current

density, $i$, as determined by a plateau in measured potential within the first 30 min of applied current, is 250 µA cm$^{-2}$. The smallest unsustainable $i$, as determined by the lack of a plateau in the potential within the first 30 min of applied current is 500 µA cm$^{-2}$. (B) Time-dependent potential behavior of 150-µm thick 1.00 M LiI-doped SN–PAN composite in response to applied current densities (50 to 750 µA cm$^{-2}$, light brown to dark brown) at 45 °C. (C) Magnification of area marked in (B). (D) Log–log plot of Sand time at various current densities at 45 °C. (E) Sand capacity as a function of applied current density at 45 °C. (F) Log–log plot of sand time at various current density for different operating temperatures: 15 °C (green), 25 °C (purple), 35 °C (blue), 45 °C (brown). (G) Temperature-dependent Sand time plot showing the Arrhenius dependence at different current densities: 50 µA cm$^{-2}$ (blue), 100 µA cm$^{-2}$ (green), 250 µA cm$^{-2}$ (brown), 500 µA cm$^{-2}$ (purple), 750 µA cm$^{-2}$ (black).

Dendrite formation is more probable when Li$^+$ is depleted at the Li–SIC interface, which signifies that the operating current is above the limiting current (*24*, *45*). The time at which the Li$^+$ concentration at the anode–electrolyte interface reaches zero denotes the Sand time and the amount of Li plated before the Sand time denotes the Sand capacity. At the Sand time, steep changes in the plating overpotential are generally observed in advance of a dendrite-related shorting of the cell. At a given current density, Sand time and Sand capacity define the safe operating limits of the system. The Sand time and Sand capacity for LiI-doped SN–PAN composite SICs each exhibit a power-law dependence on current density (Figure 6D–E), as has been observed in other polymer electrolyte systems (*25*, *46*). Above the limiting current, the slope of Sand time vs. current density in a log–log plot indicates diffusion rate for Li$^+$ in the SIC (*24*). The temperature-dependent Sand time vs. current density curve showed an increasing slope as the temperature increased (Figure 6F). This signifies a higher rate of Li$^+$ diffusion at elevated temperature, consistent with our conductivity measurements. The Sand time depends directly upon the diffusion coefficient. Direct correlation between the diffusion coefficient and Sand time means that both parameters have Arrhenius dependence on temperature as observed in our analysis (Figure 6G). Notably, it is possible to operate Li metal cells at sub-ambient temperature: e.g., 15 °C, while extracting very similar Sand capacity and time as at 25 °C (Figure 6F).

**Discussion**

Polymers are frequently paired with ion-conducting plastic crystals to render them processable and dimensionally stable for use as solid electrolytes in solid-state batteries. Our work presents a surprising and foundational molecular understanding of how polymers "dope" such composites to enhance Li$^+$ transport through the formation of a high entropy interphase between the polymer and the plastic crystal, which evidently requires specific polymers for a given choice of matrix and lithium salt. When properly configured, the composited ions partition effectively in the interphase and exhibit a higher fraction of matrix-separated ion pairs than in the bulk plastic crystal matrix. As a result, Li$^+$ diffusivity is an order of magnitude higher. Using molecular dynamics simulations, we tie the increase in Li$^+$ diffusivity to an increase in the frequency of bond rotation in the plastic crystal, which itself is tied to local increases in molar volume. With this understanding, we demonstrate that for Li halide-doped SN–PAN composites, we enable ionic conductivities of up to 5 mS cm$^{-1}$ and transference numbers up to 0.72 at 25 °C.

Until now, it has been difficult to tie macroscale transport properties in composite SICs to local Li$^+$ solid-solvation structures and matrix dynamics, which are unique to the emergent high entropy interphase. Notably, we provide evidence that while specific polymers may be necessary to induce the formation of a high entropy interphase, the polymer does not directly participate in Li$^+$ transport driving the ion current in a polarized cell. This goes against the conventional wisdom that polymers should be designed with specific molecular motifs to promote transport along polymer chains with direct participation of those functional groups in the solid-state. An alternative interpretation, which may become the vantage point for future macromolecular design, is that such functionality along the backbone should facilitate compositing and ionization at high salt loading, increase the molar volume of the plastic crystal matrix components in local proximity to the polymer to enhance local dynamics, and eventually undergo percolation to make use of faster Li$^+$ diffusivity to increase the limiting current, which enables operation of solid-state cells at ambient and sub-ambient temperatures as we have shown here for lithium halide-doped SN–PAN composites in Li metal cells.

Our use of lithium halide salts in SN–PAN composite SICs also alleviates the known incompatibility between SN and Li metal by harnessing the self-passivating nature of lithium halides at the Li–SIC interface, which affords sustainable low ASR for plating and stripping Li metal under ambient conditions at moderate current densities. These features compare quite favorably with best-in-class approaches featured elsewhere for both non-ionic and ionic plastic crystal matrices and even ceramic and polymer–ceramic composite SICs (*30*, *48*, *49*). Owing to the stability of the Li–SIC interface, we gain new insights into the dendrite-suppressing character of composite SICs in Li-metal cells and show how halide redox can be used as a means of overcharge protection (*3*, *6–8*, *22*, *23*, *27*).

These features bring to light unexpected and ultimately new opportunities for polymer-doped plastic crystal SICs to serve as solid-state electrolytes in Li metal batteries, particularly for vehicles with large powertrains where safety is paramount. The ease in which plastic crystal–polymer composite SICs may be formulated and processed in solid-state batteries indicates that there is substantially more in this design space that can be explored at the fundamental level, aided by machine learning paired with automated electrolyte formulation and screening. These data sets can further be used to train an artificial intelligence for fully autonomous co-optimization of interdependent system variables tied to the intrinsic characteristics of the composite SICs and the extrinsic characteristics when used with different cell chemistries. We anticipate in future designs that it may be possible to further tune molar volume and the frequency of bond rotation by defect engineering to enhance ion transport, as has been recently demonstrated in inorganic SICs (*50*, *51*). This could lead to a convergence in design principles for harnessing entropy and framework dynamics across materials classes to improve the prospects for solid electrolytes to enable high power solid-state batteries.

**Acknowledgements**

Funding for this work was provided by the Energy & Biosciences Institute (EBI) through the EBI–Shell program. E.J.C. was supported by an Early Career Award in the Condensed Phase and Interfacial Molecular Science Program, in the Chemical Sciences Geosciences and Biosciences Division of the Office of Basic Energy Sciences of the U.S. Department of Energy under Contract No. DE-AC02-05CH11231. Work at the Molecular Foundry—including powder X-Ray diffraction, differential scanning calorimetry, and computational studies—was supported by the Office of Science, Office of Basic Energy Sciences, of the U.S. Department of Energy under Contract No. DE-AC02-05CH11231.

## Materials and Methods

### Materials

1,3–Dioxolane (DOL), poly(acrylonitrile) (PAN, $M_w$ = 150,000 g mol$^{-1}$), lithium bromide, and lithium iodide were purchased from Sigma Aldrich. Succinonitrile (99+ % purity) was purchased from Acros Organics. Lithium metal (99.9% purity on metal basis and 1.5 mm thick) was purchased from Alfa Aesar.

### Methods

#### Preparation of Lithium Halide-Doped SN–PAN Composite SICs

Lithium halide-doped SN–PAN SICs were synthesized by dissolving lithium iodide (0.05–1 M) or lithium bromide (0.05–0.50 M) along with PAN (7.5 wt%) in molten succinonitrile at 80 °C over 24 h, after which the transparent mixtures were cooled to room temperature to solidify the SICs.

#### Preparation of Lithium Halide-Doped SN SICs

Lithium halide-doped SN electrolytes were synthesized by dissolving lithium iodide (0.05–0.50 M) or lithium bromide (0.05–0.50 M) in molten succinonitrile at 80 °C over 24 h, after which, the clear solution was cooled down to room temperature, after which the transparent mixtures were cooled to room temperature to solidify the SICs.

#### Assembly of Stainless Steel–Stainless Steel (SS–SS) Symmetric Cells

Lithium halide-doped SN–PAN composite SICs and lithium halide-doped SN SICs were each pressed inside a Chemical–Resistant PTFE Plastic Washer for Number 10 Screw Size, with 0.203" ID, 0.562" OD and thickness of 200 μm. The SSE inside the PTFE washer was then sandwiched in between two stainless steel (SS) disc shims (0.001" Thick). This setup was sealed inside a

CR2032 coin cell with a wave spring in place. The SS|SIC|SS cell was used to measure ionic conductivity.

**Preparation of Li Metal Anodes**

Lithium metal anodes were punched into 8-mm diameter discs. A solution of 0.5 M LiI in DOL (25 µl) was dropcast onto the electrode surface, after which DOL was allowed to evaporate forming a uniform layer of LiI on Li.

**Assembly of Li–Li symmetric cells**

150-µm thick sheets of lithium halide-doped SN–PAN SICs were prepared by pressing the SICs between two Teflon sheets at a pressure of 2000 psi using MTI YLJ–15 press. A SIC disk of 7/16" was punched and sandwiched in between two LiI coated Li metal discs. This setup was sealed inside a CR2032 coin cell with a wave spring in place. The Li|LiI|SIC|LiI|Li cell was used to perform strip/plate cycling, evaluate transference number, and conduct Sand analysis.

**Assembly of Cu–Li symmetric cells**

150-µm thick sheets of lithium halide-doped SN–PAN composite SICs were prepared by pressing them between two Teflon sheets at a pressure of 2000 psi using MTI YLJ–15 press. Disks (7/16") were punched and sandwiched in between 8-mm Cu disc and LiI-coated Li metal disc. This setup was sealed inside a CR2032 coin cell with a wave spring in place. The Cu|SIC|LiI|Li cell was used for linear sweep voltammetry (LSV).

**Instrumentation**

Electrochemical impedance spectroscopy (EIS) was measured at open–circuit voltage (OCV) with a Biologic VMP3 potentiostat from 1 MHz to 1 Hz with an *ac* voltage amplitude of 10 mV.

X–ray diffraction was carried out using a Bruker AXS D8 Discover GADDS X–Ray Diffractometer.

Differential Scanning Calorimetry was performed using TA Instruments Q200/RCS90 DSC at the scan rate of 1 °C min$^{-1}$ in between –80 °C and 100 °C. Samples were hermetically sealed inside the Argon glovebox.

**Computational Methodology**

To understand the electronic structure of the materials at the anode, we constructed different combinations of the materials and carried out Density Functional Theory calculations using the Vienna ab initio Simulation Package (VASP). A plane–wave basis set with a 500 eV cut–off energy and projector augmented wave (PAW) pseudopotentials were used. K-point densities were 0.05 Å$^{-1}$ and 0.3 Å$^{-1}$ for relaxation and total energy calculations, respectively. The force and energy convergence criteria were 10$^{-4}$ eV Å$^{-1}$ and 10$^{-5}$ eV Å$^{-1}$, respectively. Bader charge analysis was used to calculate net atomic charges.

We performed classical molecular dynamics (MD) simulations to understand the origin of conductivity increases with the incorporation of PAN in SN. The simulations were performed using LAMMPS software (*28*). We used the OPLS–AA force fields (*52*) to describe the interatomic interactions. For each of the simulations, a consistent standard MD procedure was carried out, which involved energy minimization, NPT MD for 5 ns and NVT MD for 20 ns for the final production trajectories. For the first system, a simulation cell containing only 3678 SN molecules was used. With OPLS-AA force fields, the simulated bulk density of the BCC crystal structure is consistent with the experimental results within 2% of error. For the second system, as shown in Figure 1(b), LiI salt with 0.10 M concentration was introduced in the first system. The formal ionic charges for Li$^+$ and I$^-$ were employed in the original OPLS-AA forcefield (+1e and –1e, respectively). However, according to our ab initio calculations for the same composite (within a smaller simulation cell), we discovered that the observed charges are closer to ±0.8e for the ions (*53*). This result is obtained using Bader charge analysis based on the ground state charge density. It is worth mentioning that employing formal charges (the ±1.0e charge scheme) resulted in strong

Coulomb attraction between the ions that could lead to unphysical precipitation or phase separation of the ions from the solvating SN, evident as large LiI clusters. Employing reduced charges (80%), consistent with the ab initio charges, did not lead to the formation of large clusters. For the third system, as shown in Figure 1(a), we introduced 7.5% (by weight of SN) PAN polymers into a crystalline SN supercell and distributed 0.10 M LiI as before. For the last 10 ns of the MD simulations, we calculated the Mean Square Displacement (MSD) of the Li$^+$ ions in the systems shown in Figure 1 (a) and (b), the results are shown in Figure 1(c). The slope of MSD vs time provides the diffusion coefficient ($D$) of Li$^+$ ions via the Einstein relation.

**Supplementary Information**

**Supplementary Figures**

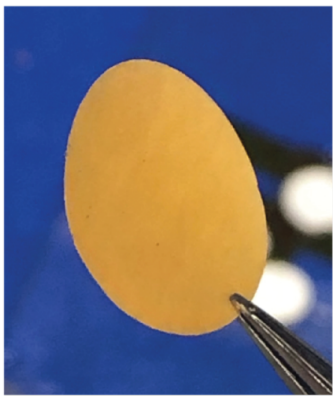

**Figure S1** Image of 150-μm thick disk of 1.0 M LiI-doped SN–PAN composite solid-ion conductor (SIC).

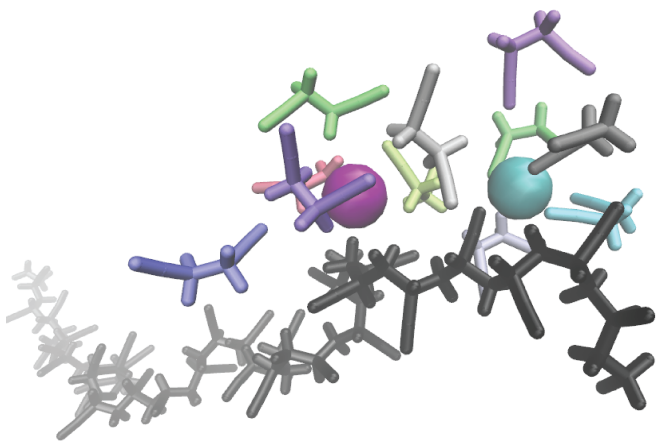

**Figure S2** Close up look at the PAN–SN interface inside the relaxed SIC simulation cell. The PAN (black), and the SN (different colored for clarity) in the vicinity of the Li$^+$ (cyan) and I$^-$ (purple). The longer segments pointing out of PAN backbone and on SN are the nitrile group, where the shorter ones are hydrogen.

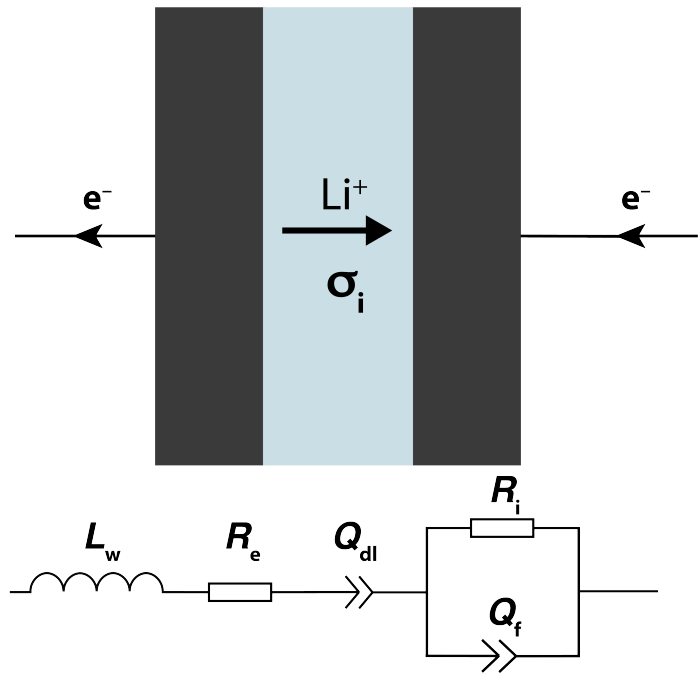

**Figure S3** Schematic of the cell consisting of 150-μm thick SIC film assembled between stainless-steel (SS) ion-blocking electrodes (top), which was used to perform electrochemical impedance spectroscopy (EIS). The Nyquist plot was fitted with the equivalent circuit (bottom) to obtain the ion conductivity of the SIC. Each element of the equivalent circuit represents a physical process undergoing within the cell during the EIS measurement. $L_w$ and $R_e$ represent inductance and electronic resistance of the electrical cable between the cell and the power source (respectively), $Q_{dl}$ represents double layer capacitance due to the charge storage at SS and SIC interfaces, and $Q_f$ and $R_i$ represent bulk capacitance and ionic resistance of the SIC film (respectively). The ionic conductivity of the film, $\sigma_i$, equals $\dfrac{L}{R_i A}$, where $L$ is the thickness of the SIC film, $A$ is the projected area, and $R_i$ is the fitted ionic resistance of the SIC film.

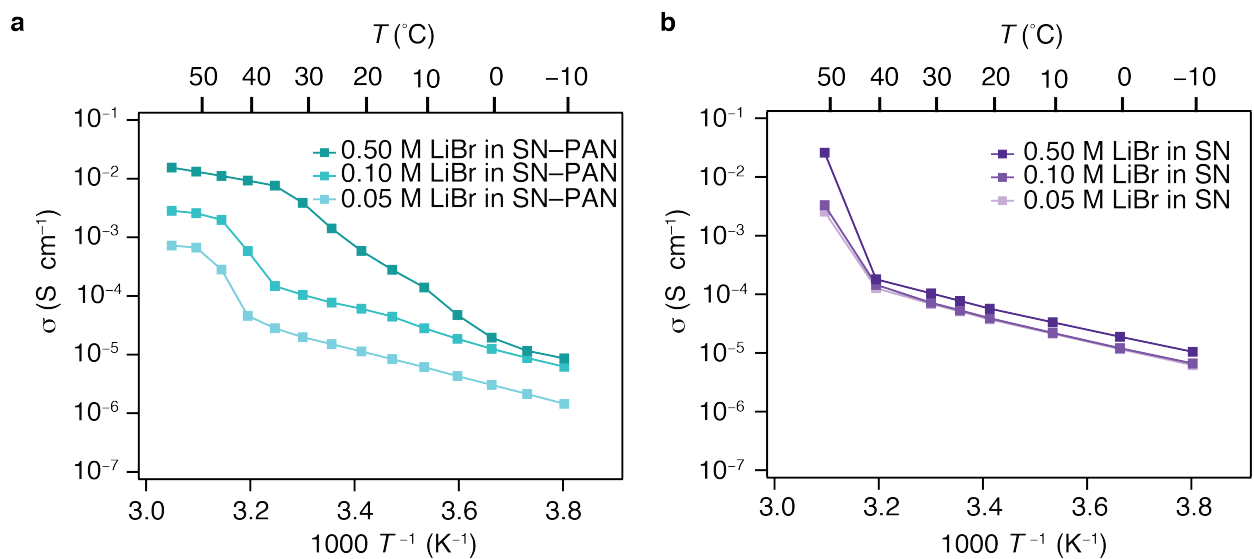

**Figure S4** (A) Temperature-dependent ionic conductivity of SN–PAN composite doped with [LiBr] = 0.05 M, 0.10 M, and 0.50 M (light cyan to dark cyan). (B) Temperature-dependent ionic conductivity of SN doped with [LiBr] = 0.05 M, 0.10 M, and 0.50 M (light purple to dark purple).

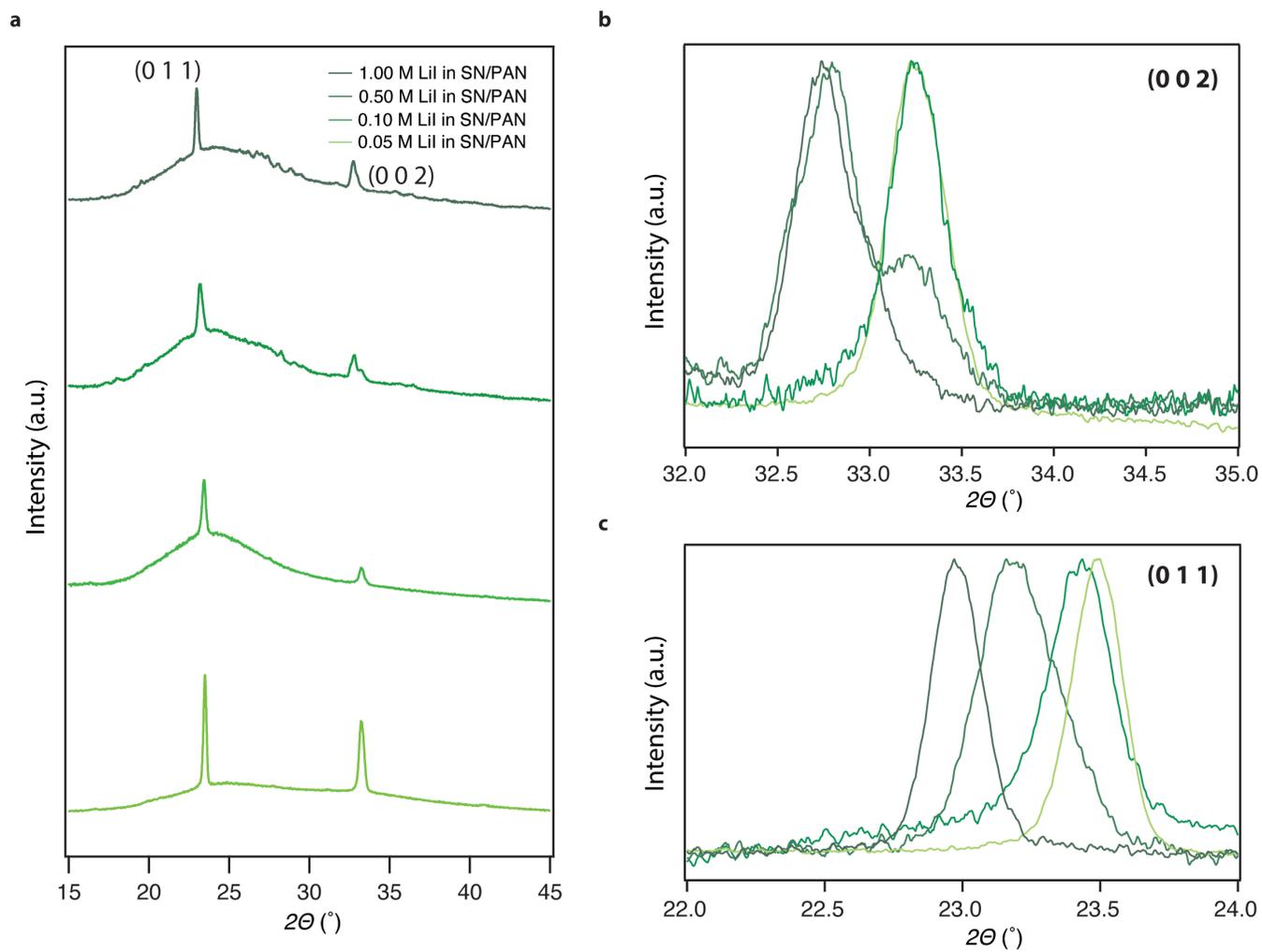

**Figure S5** (A-C) X–Ray diffraction (XRD) patterns of SN–PAN composites doped with [LiI] = 0.05 M, 0.10 M, 0.50 M, and 1.00 M (light green to dark green).

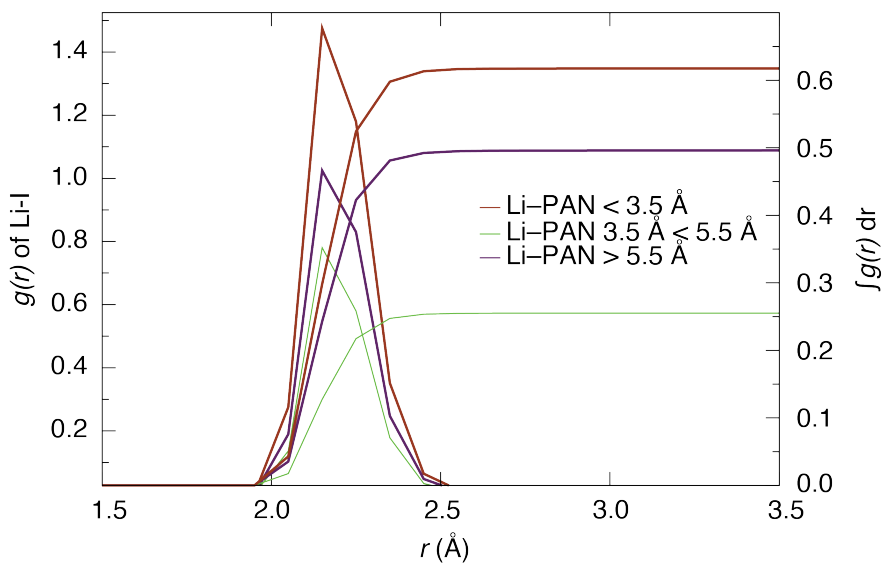

**Figure S6** Radial distribution function, $g(r)$, of Li$^+$ concentration with respect to the distance from I$^-$ within three domains in SIC—within 3.5 Å (brown zone), between 3.5 Å and 5.5 Å (green zone), and more than 5.5 Å (purple zone) from the PAN backbone.

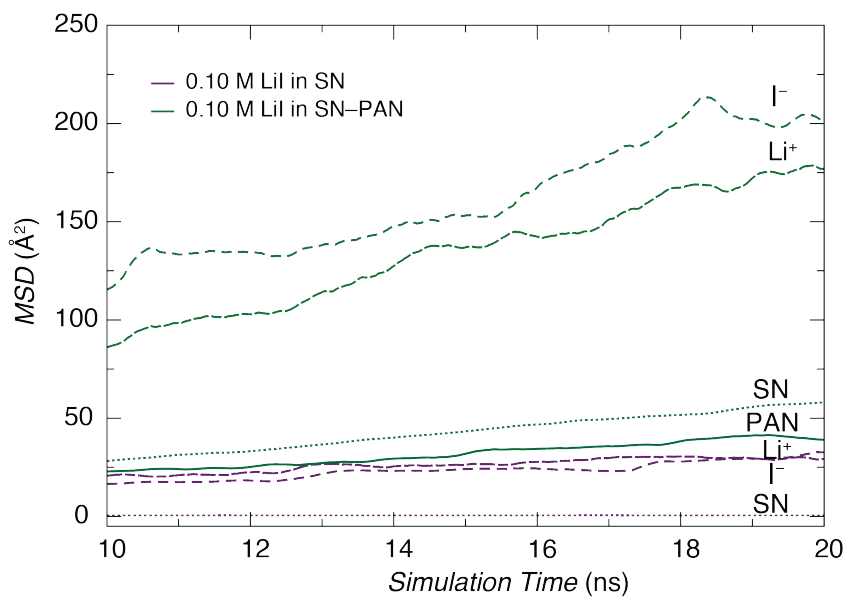

**Figure S7** The simulated mean square displacement (*MSD*) of Li$^+$, SN molecules and I$^-$ as a function of simulation time within 0.10 M LiI-doped SN–PAN composite SIC (green) and 0.10 M LiI-doped SN SIC (purple). It is evident from the *MSD* that translation motion of SN molecules is negligible in comparison to Li$^+$ and I$^-$ ions within the SIC.

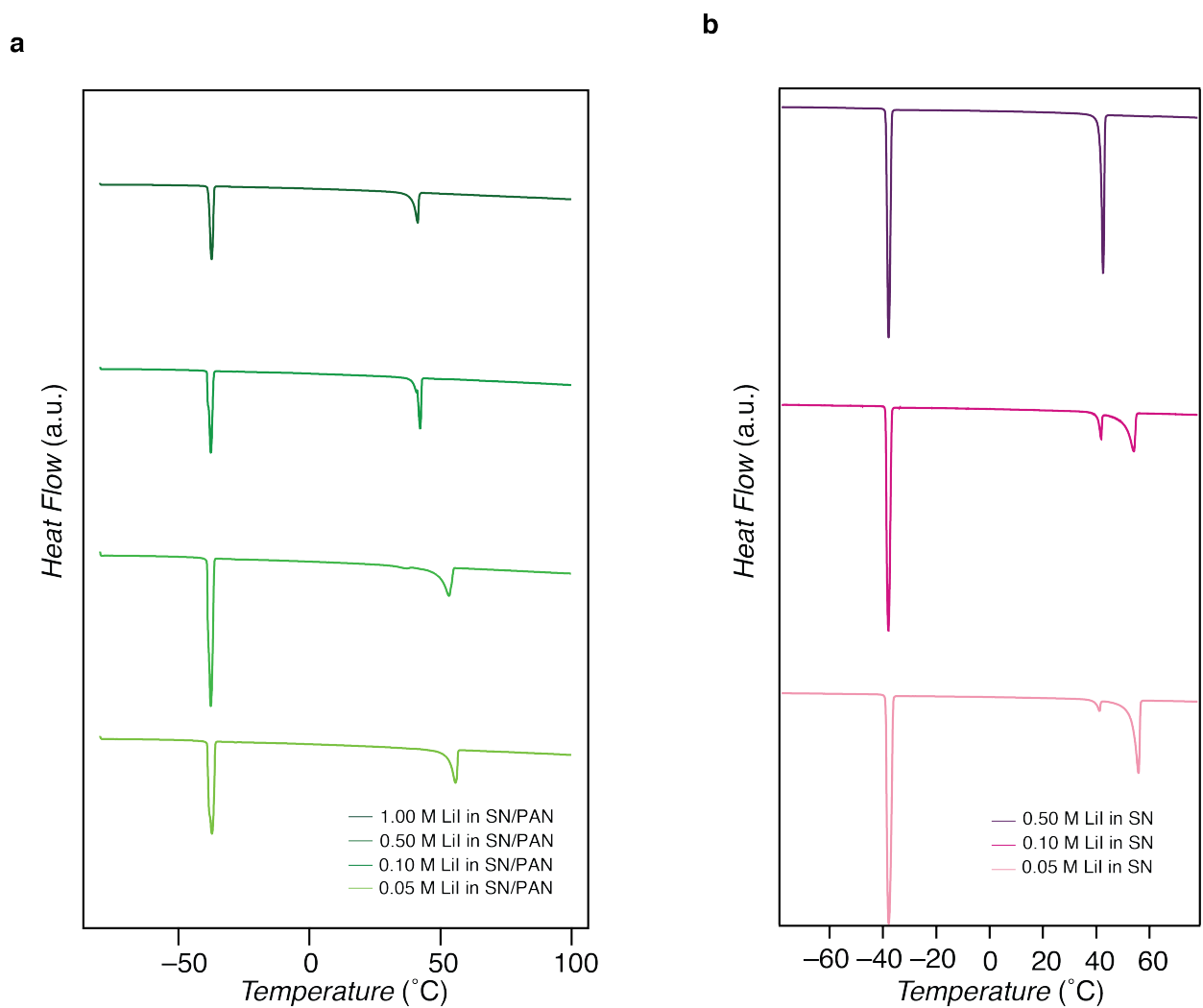

**Figure S8** (A) Differential scanning calorimetry (DSC) SN–PAN composites doped with [LiI] = 0.05 M, 0.10 M, 0.50 M, and 1.00 M (light green to dark green). (B) DSC of SN doped with [LiI] = 0.05 M, 0.10 M, and 0.50 M (light purple to dark purple).

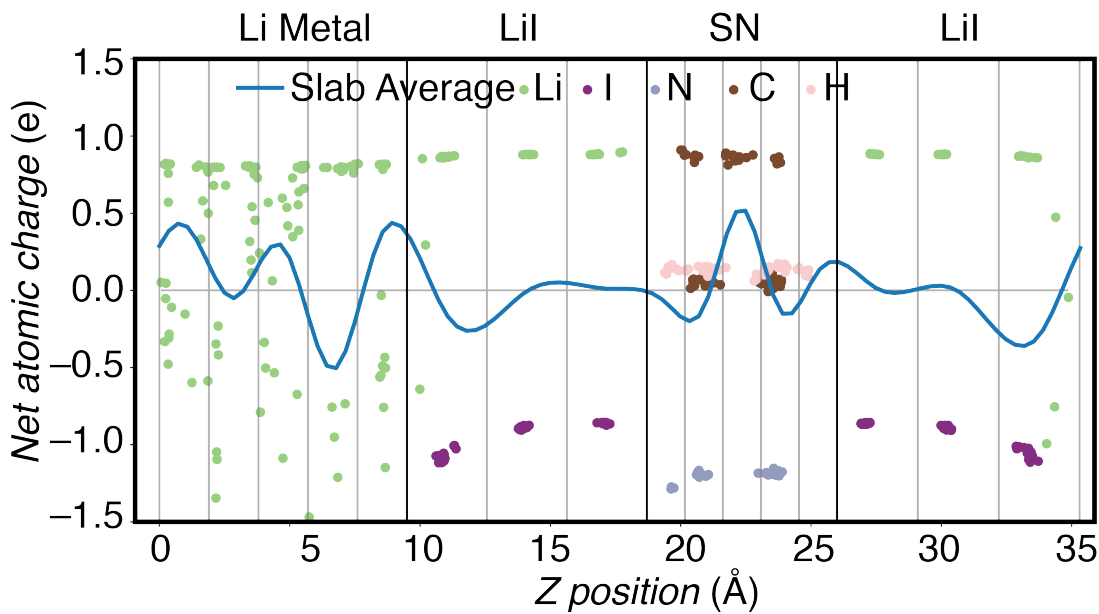

**Figure S9** The scatter plots are the simulated net charge on each atom which is labeled on the y-axis and plotted against the z coordinates of the atom. The blue line is the average charge of the atoms inside each bin defined by the vertical grey lines. At the Li metal and LiI interface, a clear electrical double layer is formed having metal side positively charged and the halide negatively charged, and the degree of charge for the interfacial I ions are larger than that in the bulk part of LiI, this indicates charge transferred from metal to LiI crystal.

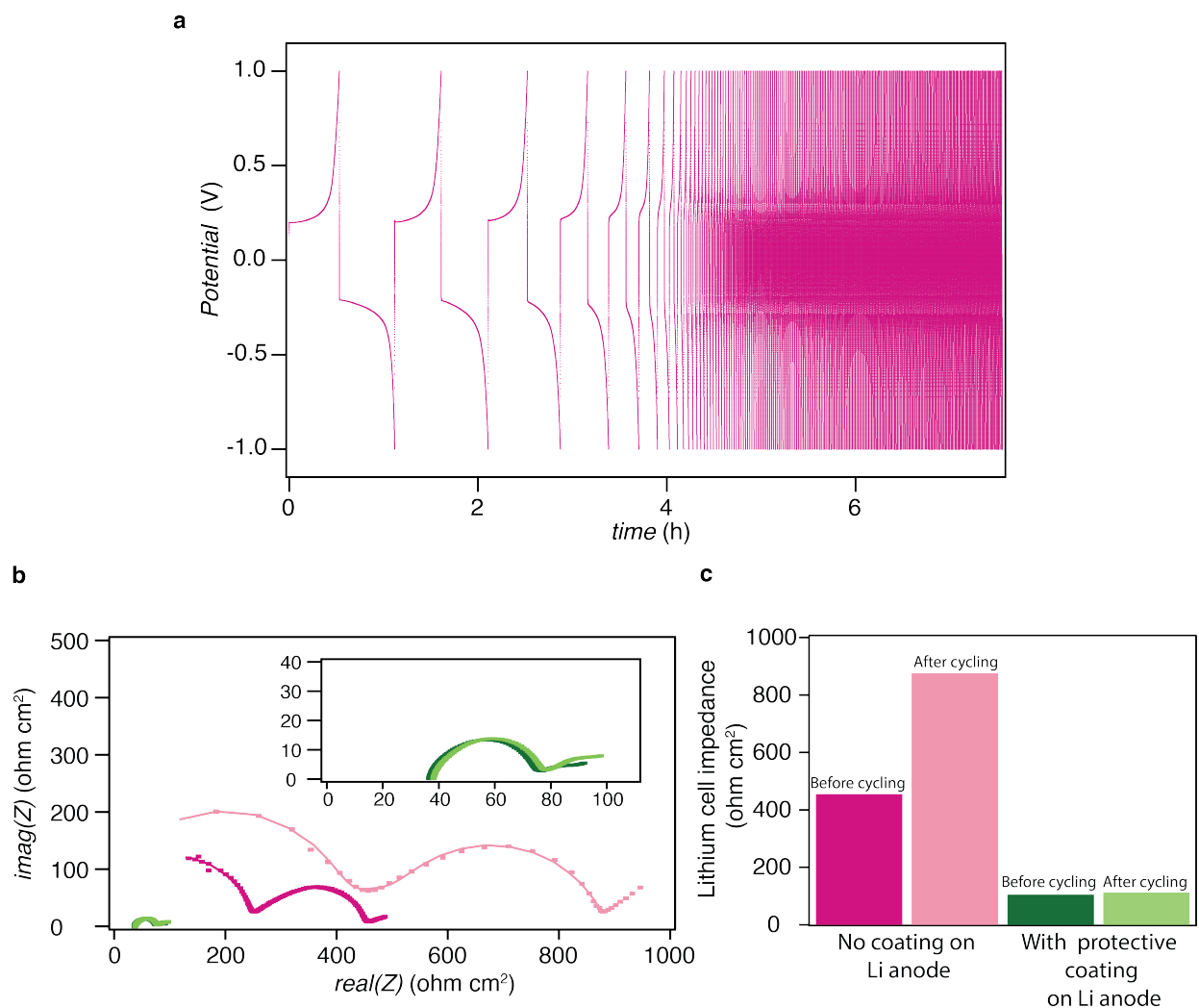

**Figure S10** (A) Reversible lithium plating in 0.50 M LiI-doped SN SIC sandwiched between two lithium disks at a current density of 25 µA cm$^{-2}$. Direct contact between Li metal and SN electrolyte leads to the formation of insulating SEI layer. Insulating layer substantially increases the overpotential required to plate/strip lithium. The growth of an insulating SEI over time leads to increase in plating overpotential and shortens the operating time. (B) Impedance spectroscopy of Li–Li symmetric cells with (green) and without (pink) LiI at the Li metal anode interface. Light colors indicate before any cycling and dark colors indicate after 100 cycles of plating and stripping. (C) Extracted cell impedance before and after 100 cycles of plating and striping for Li–Li symmetric cells, with and without LiI on Li metal.